\RequirePackage{pdf14}
\documentclass[reprint,
	amsmath,amssymb,
	aps,
	pra,
	floatfix,
]{revtex4-2}

\usepackage{graphicx}\usepackage{dcolumn}\usepackage{bm}\usepackage{hyperref}\usepackage{svg}
	\usepackage{braket}
	\usepackage{mathtools}
	\usepackage{color}
	\usepackage{orcidlink}
	\usepackage{dsfont}
	\usepackage{overpic}
	\usepackage{siunitx}

\usepackage{soul}

\hypersetup{
 colorlinks = true,
 linkcolor = {blue},
 citecolor = {blue},
 urlcolor = {blue},
}

\usepackage[whole]{bxcjkjatype}
\usepackage[normalem]{ulem}
\usepackage{soul}

\begin{document}

\title{Orthogonal frequency-division multiplexing for simultaneous gate operations \\on multiple qubits via a shared control line}

\author{Haruki Mitarai\orcidlink{0000-0002-8374-3648}}
 \email{hmitarai@mosk.tytlabs.co.jp}\author{Yukihiro Tadokoro\orcidlink{0000-0002-1844-3040}}\email{y.tadokoro@ieee.org}
\author{Hiroya Tanaka\orcidlink{0000-0002-5113-0927}}\email{tanak@mosk.tytlabs.co.jp}

\affiliation{Toyota Central R\&D Labs., Inc., 41-1, Yokomichi, Nagakute, Aichi 480-1192, Japan}

\date{\today}

\begin{abstract}
	The increasing number of qubits in quantum processors necessitates a corresponding increase in the number of control lines between the processor, which is typically operated at cryogenic temperatures, and external electronics.
Scaling poses significant challenges in terms of 
the thermal loads, forming a major bottleneck in the realization of large-scale quantum computers. 
In this study, we analyze simultaneous gate operations on multiple qubits using microwaves transmitted via a single cable in a frequency-division multiplexing (FDM) scheme. 
By employing rectangular control microwave pulses, we reveal the contribution of drive frequency spacing to gate fidelity.  
	Through theoretical and numerical analyses, we demonstrate that orthogonal and quasi-orthogonal microwave signals suppress interference in simultaneously driven qubits, thereby ensuring high gate fidelity. 
	Additionally, we provide design guidelines for key parameters, including pulse length, number of multiplexed microwave signals, and rotation angle, to achieve precise qubit operations.
	Our findings enable a scalable FDM-based microwave control scheme suitable for quantum processors with a large number of qubits.

\end{abstract}

\maketitle

\section{Introduction}
	Quantum computing is expected to solve problems intractable to classical computers within realistic timescales~\cite{nielsen2010quantum}. 
	The potential applications include prime factorization~\cite{Shor1}, 
	quantum chemistry calculations~\cite{kandala2017hardware, Hartree-Fock, tomaru2024}, 
	quantum and classical system simulations~\cite{Universal_Quantum_Simulators, Ising_nature_2023, PRA_wave_equation, sato_CAE}, 
and others~\cite{Grover1, Harrow, VQE, QAOA, QSVM, QCL, gravitational, ishida2025}.
	These capabilities have attracted significant attention from both academic and industrial research communities.
	In recent years, substantial research and engineering efforts have been directed toward the development of fault-tolerant quantum computers~\cite{nakamura1999coherent, blais2004cavity, Charge-insensitive, demonstration, IBM_qubit, PhysRevLett.129.030501, bluvstein2024logical, PRXQuantum.5.030347, optical_QC}. 
	For example, quantum supremacy and error correction using surface codes have been experimentally demonstrated using superconducting quantum processors~\cite{google, google2025quantum}.

	Classical control electronics are crucial for the advancement of quantum processors because they strongly affect the precision of qubit manipulation~\cite{Impact_of}. 
	Currently, the control architectures for many types of qubits commonly generate microwave signals at ambient temperatures~\cite{GaAs_qubit, refrigerator, VANDIJK201990,  Si_qubit_cmos, takeuchi2024microwave}. 
	The microwave signals are then delivered to individual qubits, typically located in a dilution refrigerator, via one-to-one coaxial cabling.
	However, this approach faces scalability challenges for large-scale quantum computers.  
	As the number of qubits increases, the number of required control lines increases. 
	Typically, exploiting quantum advantages for solving realistic problems requires millions of qubits~\cite{Surface_codes, gameofsurfacecodes}. 
Note that increasing the number of control wires not only adds hardware complexity and cost but also imposes additional heat loads through physical interconnections, thereby threatening the cooling capacity of the refrigerators~\cite{refrigerator}. 

Hence, wire reduction is a critical focus in scaling quantum computing systems~\cite{IEEE_cryo_2019, Cryo-CMOS_FDMA, PhysRevApplied.18.064046, Baseband, PhysRevResearch.5.013145, IOP_cryo_2023, ohira2024optimizing, Roberto_May2025}. 
Frequency-division multiplexing (FDM) is an essential technique that enables a single microwave cable to simultaneously control multiple qubits, thereby significantly reducing the number of required control lines.
	However, a microwave tone intended for one qubit is off-resonant for, and interferes with, other simultaneously driven qubits, thus hindering precise parallel gate operations~\cite{Impact_of, ohira2024optimizing}.
Therefore, suppressing such interference is essential to achieve high gate fidelity in the FDM scheme. 
	Accordingly, a frequency-multiplexed qubit controller, based on adiabatic quantum-flux-parametron logic and a superconducting resonator array operating as a frequency demultiplexer, was investigated~\cite{takeuchi2024microwave}.
In addition, selective gate operations using pulse shaping have been demonstrated~\cite{selective_excitation}. 

In this study, we investigate simultaneous gate operations on multiple qubits driven by a frequency-division-multiplexed microwave signal. 
	We focus on the role of drive frequency spacing and its impact on gate fidelity.
	Through theoretical and numerical analyses, we demonstrate that orthogonal FDM (OFDM)~\cite{OFDM, OFDM2009, 5G} can mitigate interference from off-resonant microwave components 
owing to their orthogonality, resulting in high gate fidelity.
	In addition, at half the frequency spacing required for pulse orthogonality, the gate fidelity is comparable to that achieved with orthogonal spacing, even though the multiplexed microwaves are no longer mutually orthogonal.
	Furthermore, we discuss the design requirements for the key parameters, including pulse length, number of multiplexed microwave signals, and rotation angle, to achieve high-fidelity gate operations. 

The developed framework enables the use of closely spaced drive frequencies and short pulse lengths, thereby achieving high throughput in gate operations. 
	Our approach is inherently scalable and well suited for controlling a large number of qubits owing to its simplicity and narrow bandwidth requirements.

\section{Model} \label{sec:model}
	\begin{figure}[t]
		\centering
		\includegraphics[width = 3.4in]{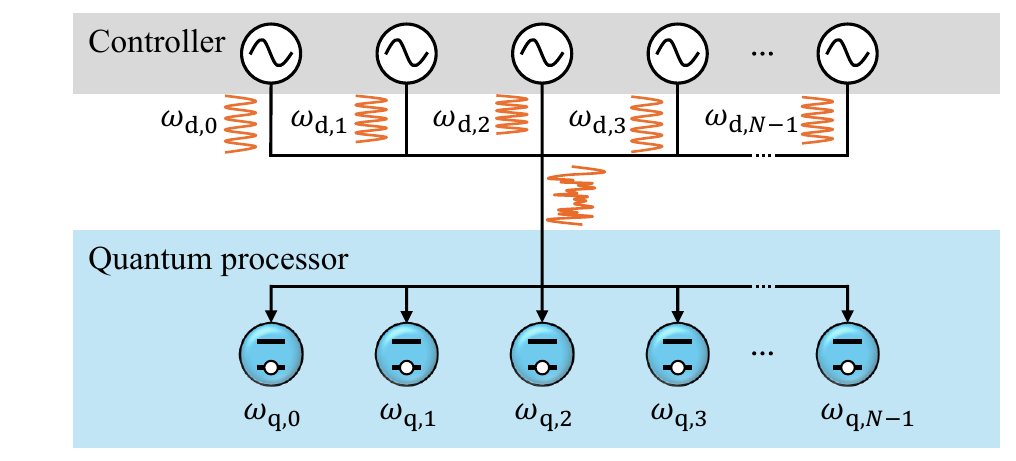}
		\caption{Conceptual illustration of a model comprising $N$ independent qubits with frequencies $\omega_{\mathrm{q}, k}$ driven by microwaves via a shared control line. 
		The controller produces $N$ microwave tones at distinct frequencies $\omega_{\mathrm{d}, k}$ with arbitrary envelopes, which are then combined and routed to a quantum processor via the shared line. 
		The combined signal is applied uniformly to all qubits. 
		}
		\label{fig:model}
	\end{figure}

	Figure~\ref{fig:model} shows a system of $N$ independent qubits driven by a frequency-division-multiplexed microwave signal.
	The controller generates microwaves with $N$ distinct frequencies that are then combined and transmitted to the quantum processor through a shared control line. 
	The synthesized signal is applied uniformly to all the qubits.
	The Hamiltonian of the entire system is given by 
	\begin{align}
		\label{eq:H_all}
		&H_{\text{all}}\left(t\right) \nonumber \\ 
		&\quad\equiv \sum_{k \in K}\left[ -\frac{\omega_{\mathrm{q}, k}}{2} \sigma_{\mathrm{z}, k} + 
		\sum_{j \in K} \alpha_j s\left(t\right) \sin\left(\omega_{\mathrm{d}, j} t + \theta_j \right) \sigma_{\mathrm{y}, k}\right], 
	\end{align}
	where $\omega_{\mathrm{d}, k}$, $\theta_k$, and $\alpha_k$ denote the frequency, phase, and amplitude, respectively, of the microwave driving the $k$th qubit, which has a resonance frequency $\omega_{\mathrm{q}, k}$. $K \subset \mathbb{Z}$ is the set of qubit indices with cardinality $\left| K \right| = N$. 
	The operators $\sigma_{\mathrm{z}, k}$ and $\sigma_{\mathrm{y}, k}$ are the Pauli $Z$ and $Y$ operators associated with the $k$th qubit. 
	The function $s\left(t\right)$ represents the pulse envelope. 
	Here, we set $\hbar = 1$ for simplicity.

	Since arbitrary single-qubit gates can be implemented with $X$ and virtual-$Z$ gates~\cite{virtual_Z}, we consider only the driving terms involving $\sigma_{\mathrm{y}, k}$.
	Now, we focus on the qubit $k = k_0$ in Eq.~\eqref{eq:H_all} and analyze the system using the single-qubit Hamiltonian given by 
	\begin{equation}
		\label{eq:H}
		H \left(t\right)\equiv -\frac{\omega_{\mathrm{q}, k_0}}{2} \sigma_{\mathrm{z}, k_0} + \sum_{k \in K}  \alpha_k s\left(t\right) \sin\left(\omega_{\mathrm{d}, k} t + \theta_k \right) \sigma_{\mathrm{y}, k_0}, 
	\end{equation}
	where $k_0 \in K$ is the index of the target qubit.
	Equation~\eqref{eq:H} describes the dynamics of the $k_0$th qubit under a multiplexed microwave drive.
	
	We then move to the frame rotating at the qubit frequency $\omega_{\mathrm{q}, k_0}$. 
	Let $\Ket{\psi_{\mathrm{I}}\left(t\right)} \equiv e^{iH_0t} \Ket{\psi\left(t\right)}$ be the state vector in the rotating frame, 
	where $H_0 \equiv -\frac{\omega_{\mathrm{q}, k_0}}{2} \sigma_{\mathrm{z}, k_0}$
	and $\Ket{\psi\left(t\right)}$ denotes the state vector in the laboratory frame.
	The time evolution of the system is described by the Schr\"{o}dinger equation in the frame rotating at $\omega_{\mathrm{q}, k_0}$: 
	$i \dot{\Ket{\psi_{\mathrm{I}}}} = H_{\mathrm{I}} \Ket{\psi_{\mathrm{I}}}$, where
	\begin{align}
		\label{eq:HI}
		&H_{\mathrm{I}} \left(t\right) = \frac{1}{2} s\left(t\right) \nonumber\\
		&\quad\sum_{k \in K}  \alpha_k\left[
			\sigma_{\mathrm{x}, k_0} \left\{ \cos \left(\left(\omega_{\mathrm{d}, k} + \omega_{\mathrm{q}, k_0}\right)t\right) - \cos \left(\left(\omega_{\mathrm{d}, k} - \omega_{\mathrm{q}, k_0}\right) t\right)\right\} \right.\nonumber\\
			&\quad\left.+ \sigma_{\mathrm{y}, k_0} \left\{\sin \left(\left(\omega_{\mathrm{d}, k} + \omega_{\mathrm{q}, k_0}\right)t\right) + \sin \left(\left(\omega_{\mathrm{d}, k} - \omega_{\mathrm{q}, k_0}\right) t\right)\right\}
		\right].
	\end{align}
	Here, $\sigma_{\mathrm{x}, k}$ represents the Pauli $X$ operator associated with the $k$th qubit.
	For simplicity, we assume $\theta_k = 0$ throughout this paper.

	We assume that the microwave pulses have rectangular envelopes. 
	Additionally, we set $\omega_{\mathrm{q}, k} = \omega_{\mathrm{d}, k}$ for all $k \in K$, and allocate $\omega_{\mathrm{q}, k}$ (and $\omega_{\mathrm{d}, k}$) at regular intervals $\Delta \ll \omega_{\mathrm{q}, k}, \omega_{\mathrm{d}, k}$. 
	We apply the same gate operation to all the qubits simultaneously. 
	Therefore, we set $\alpha_k = \alpha$ for all $k$.
	Under these assumptions, the system Hamiltonian $H_{\mathrm{I}}$ in Eq.~\eqref{eq:HI} is expressed as 
	\begin{align}
		\label{eq:HI2}
		H_{\mathrm{I}} \left(t\right) = & \frac{\alpha}{2} s\left(t\right)
		\nonumber \\ &
			\sum_{k \in K}  \left[
			\sigma_{\mathrm{x}, k_0} \left\{\cos 
			\left( \left(2\omega_{\mathrm{q}, k_0} + \Delta_{k k_0} \right)t\right) - \cos \left(\Delta_{k k_0} t \right)\right\} \right.\nonumber\\
			&\left.+ \sigma_{\mathrm{y}, k_0} \left\{\sin \left(\left(2\omega_{\mathrm{q}, k_0} + \Delta_{k k_0}\right)t\right) + \sin \left( \Delta_{k k_0} t \right)\right\}
		\right] ,
	\end{align}
	where $\Delta_{kj} \equiv (k-j)\Delta$, 
	\begin{align}
		\label{eq:s}
		s\left(t\right) = \left\{ \,
			\begin{aligned}
				&1 \quad 0 \le t \le \tau\\
				&0 \quad \text{otherwise}
			\end{aligned}
			\right. , 
	\end{align}
	and $\tau$ is the pulse length. 
	In this context, the qubit index set $K$ is given by $K = \left\{ l, l+1, \dots , r-1, r\right\}$, where $r - l + 1 = N$.
    The integers $l \le 0$ and $r \ge 0$ denote the left and right boundaries of the index set.

\section{General analysis} \label{sec:general_FDM}
	\begin{figure}[t]
			\centering
			\includegraphics[width = 3.4in]{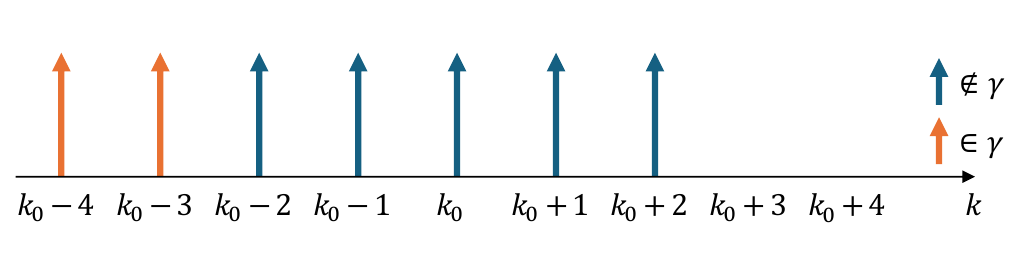}
		\caption{Illustration of $\gamma$, which represents the set of qubit indices lacking a corresponding element with respect to $k_0$. 
		The orange and blue arrows denote the driving microwaves that correspond and do not correspond to the set $\gamma$, respectively.
		In this figure, the set $\gamma$ is specifically $\left\{ k_0-3,k_0-4 \right\}$. These indices have no counterpart related to $k_0$, i.e., there are no arrows at $k_0+3$ and $k_0+4$.
}
		\label{fig:gamma}
	\end{figure}
In FDM-based systems, the microwave component driving a specific qubit interferes with the operation of other qubits. 
	Hence, suppressing the interference from the off-resonant microwave components is crucial. 
To quantify the interference from off-resonant microwaves, we use the average gate fidelity $F\left(U_1, U_2\right)$, defined as~\cite{fidelity}
	\begin{align}
		\label{eq:average_gate_fidelity}
		F\left(U_1, U_2\right) \equiv \frac{\left| \text{Tr}\left(U_{1}^{\dagger}U_2\right)\right|^2 + d}{d\left(d + 1\right)}, 
	\end{align}
	where $U_{1}$ and $U_{2}$ are unitary operators and $d$ is the dimension of the Hilbert space.
    Since $d =2$, the fidelity satisfies $1/3 \le F\left(U_1, U_2\right) \le 1$. 
	Ideally, quantum computers should have high average gate fidelity $F\left(U_{\text{ideal}}, U\right)$ while simultaneously realizing closely spaced drive frequencies or short pulse lengths.
	In the remainder of this paper, $U_{\text{ideal}} \equiv e ^ {i\frac{\phi}{2} \sigma_{\mathrm{x}, k_0}}$ denotes the desired unitary operator on a two-dimensional Hilbert space with the rotation angle $\phi$. 
	The operator $U$ is defined as $\Ket{\psi_{\mathrm{I}}\left(\tau\right)} = U\Ket{\psi_{\mathrm{I}}\left(0\right)}$.

	We employ the second-order Magnus expansion~\cite{Magnus} to approximate the evolution of the state $\Ket{\psi_{\mathrm{I}}}$. 
	We define the approximate evolution operator $U_{\text{Magnus}}$ as 
	\begin{align}
		\label{eq:Magnus}
		U_{\text{Magnus}}\left(t\right) \equiv \exp\left[-i\left\{\Omega_1\left(t\right) + \Omega_2\left(t\right)\right\}\right], 
	\end{align}
	where
	\begin{align}
		\label{eq:Omega_1}
		\Omega_1\left(t\right) &\equiv \int_{0}^{t} H_{\mathrm{I}}\left(t_1\right) \mathrm{d}t_1 , 
	\end{align}
	\begin{align}
		\label{eq:Omega_2}
		\Omega_2\left(t\right) &\equiv -\frac{i}{2} \int_{0}^{t} \int_{0}^{t_1} \left[ H_{\mathrm{I}}\left(t_1\right), H_{\mathrm{I}}\left(t_2\right) \right] \mathrm{d}t_2 \mathrm{d}t_1.
	\end{align}
	To obtain the closed-form expressions for Eqs.~\eqref{eq:Omega_1} and \eqref{eq:Omega_2}, 
	we approximate $H_{\mathrm{I}}$ as 
	\begin{align}
		\label{eq:HI_RWA}
		H_{\mathrm{I}} \left(t\right)
		 & \sim \frac{\alpha}{2} s \left(t\right) \nonumber \\ & \sum_{k \in K} \left[
		- \sigma_{\mathrm{x}, k_0} \cos \left( \Delta_{kk_0} t \right)
		+ \sigma_{\mathrm{y}, k_0} \sin \left( \Delta_{kk_0} t \right) \right],
	\end{align}
	by neglecting the fast oscillating terms in Eq.~\eqref{eq:HI2}. Substituting Eq.~\eqref{eq:HI_RWA} instead of Eq.~\eqref{eq:HI2} into Eq.~\eqref{eq:Omega_1}, we obtain
	\begin{align}
		\label{eq:Omega_1_1}
		\Omega_1\left(\tau\right) 
\sim - \left( \sigma_{\mathrm{x}, k_0} \lambda_{\mathrm{x}} +  \sigma_{\mathrm{y}, k_0} \lambda_{\mathrm{y}}\right), 
	\end{align}
	where
	\begin{align}
		\label{eq:lambda_x_general}
		\lambda_{\mathrm{x}} &\equiv  
		\frac{\alpha\tau}{2}  \left[ 1 +  \sum_{\substack{k \in K \\ k \ne k_0}} \mathrm{sinc} (\Delta_{kk_0} \tau) \right],
	\end{align}
	\begin{align}
		\label{eq:lambda_y_general}
		\lambda_{\mathrm{y}} &\equiv  -
		\frac{\alpha \tau}{2} 
		\sum_{k \in \gamma} \sin \left(\frac{\Delta_{kk_0} \tau}{2}\right) 
		\mathrm{sinc} \left(\frac{\Delta_{kk_0} \tau}{2}\right) 
		,
	\end{align}
	by using $\mathrm{sinc}\left(x\right) \equiv \sin\left(x\right)/x$.
	Here, $\gamma \subset K$ is the set of qubit indices that lack corresponding elements related to $k_0$. 
    For example, let $k_0 = 0$. If $-2 \in K$ while $2 \notin K$, then $-2 \in \gamma$.
    More formally, a qubit index $k^\prime \in K$ belongs to $\gamma$ if and only if its symmetric counterpart with respect to $k_0$, namely, $ -\left(k^\prime - k_0 \right) + k_0$, does not belong to $K$ (see Fig.~\ref{fig:gamma}). 

Substituting Eq.~\eqref{eq:HI_RWA} into Eq.~\eqref{eq:Omega_2}, we obtain
	\begin{align}
		\label{eq:Omega_2_1}
		\Omega_2\left(\tau\right) \sim  - \sigma_{\mathrm{z}, k_0} \lambda_{\mathrm{z}}, 
	\end{align}
	where 
\begin{widetext}
	\begin{align}
		\label{eq:lambda_z_general}
		\lambda_{\mathrm{z}} &\equiv
		 \frac{\alpha^2 \tau}{4} 
		\left[ \sum_{k \in \gamma} \frac{ \cos(\Delta_{kk_0} \tau ) -  \frac{3}{2}\mathrm{sinc}(\Delta_{kk_0} \tau)}{\Delta_{kk_0}}
+ \sum_{k \in \gamma}
		\sum_{\substack{j \in K \\ j \ne k_0}} \left\{ 
		\frac{\mathrm{sinc}\left(\Delta_{jk_0}\tau \right)}{\Delta_{kk_0}} + \frac{\mathrm{sinc}\left[\left(\Delta_{kk_0} + \Delta_{jk_0}\right)\tau\right]}{2\Delta_{jk_0}}
		\right\} \right. \nonumber \\
		&\qquad\qquad\qquad\qquad\left.- \sum_{k \in K}
		\sum_{j \in \gamma} \frac{\mathrm{sinc}\left[\left(\Delta_{kk_0} + \Delta_{jk_0}\right)\tau\right]}{2\Delta_{jk_0}}
- \sum_{k \in \gamma}
		\sum_{\substack{j \in K \\ j \ne k_0 \\ j \ne k}} \left\{ \frac{1}{\Delta_{kk_0}} + \frac{1}{\Delta_{jk_0}} \right\}\frac{\mathrm{sinc}(\Delta_{kj}\tau)}{2}\right]. 
	\end{align}
	\end{widetext}
	Inserting Eqs.~\eqref{eq:Omega_1_1} and \eqref{eq:Omega_2_1} into Eq.~\eqref{eq:Magnus}, we obtain
	\begin{align}
		\label{eq:Magnus_1}
		& U_{\text{Magnus}}\left(\tau\right) \nonumber \\ 
		& \, \sim  \mathbb{I} \cos\left(\Lambda\right) 
		+ i \frac{1}{\Lambda} \left(\lambda_{\mathrm{x}} \sigma_{\mathrm{x}, k_0} + \lambda_{\mathrm{y}} \sigma_{\mathrm{y}, k_0} + \lambda_{\mathrm{z}}\sigma_{\mathrm{z}, k_0}\right)\sin\left(\Lambda\right), 
	\end{align}
	where 
	\begin{align}
		\label{eq:Lambda}
		\Lambda \equiv \sqrt{\lambda_{\mathrm{x}}^2 + \lambda_{\mathrm{y}}^2 + \lambda_{\mathrm{z}}^2}, 
	\end{align}
	and $\mathbb{I}$ is the identity operator. 
	From Eq.~\eqref{eq:Magnus_1}, the average gate fidelity $F\left(U_{\text{ideal}}, U_{\text{Magnus}}\left(\tau\right)\right)$ can be expressed as 
	\begin{align}
		\label{eq:F_analytical}
		&F\left(U_{\text{ideal}}, U_{\text{Magnus}}\left(\tau\right)\right) \nonumber \\
		&\; \sim \frac{1}{3} \left( 2 \left| \cos\left|\frac{\phi}{2}\right| \cos\Lambda + \frac{\phi \lambda_{\mathrm{x}}}{\left|\phi\right|\Lambda} \sin\left|\frac{\phi}{2}\right| \sin \Lambda \right|^2 + 1 \right). 
	\end{align}
	Equation~\eqref{eq:F_analytical} indicates that high fidelity, $F\left(U_{\text{ideal}}, U_{\text{Magnus}}\left(\tau\right)\right) \sim 1$, can be attained in regimes where $\lambda_{\mathrm{x}} \sim \phi/2$ and $\lambda_{\mathrm{y}}, \lambda_{\mathrm{z}} \sim 0$. 

\section{Case studies} \label{sec:case_study_FDM}
	\begin{figure}[t]
		\centering
			\begin{minipage}[t]{1\linewidth}
				\centering
				\begin{overpic}[scale=1]{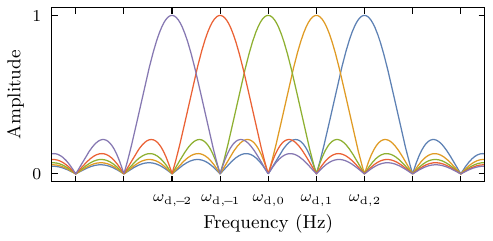}
					\put(12, 42.5){(a)}
\end{overpic}
			\end{minipage}
			\begin{minipage}[t]{1\linewidth}
				\centering
				\begin{overpic}[scale=1]{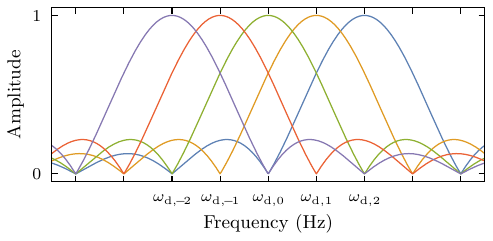}
					\put(12, 42.5){(b)}
				\end{overpic}
			\end{minipage}
			\begin{minipage}[t]{1\linewidth}
				\centering
				\begin{overpic}[scale=1]{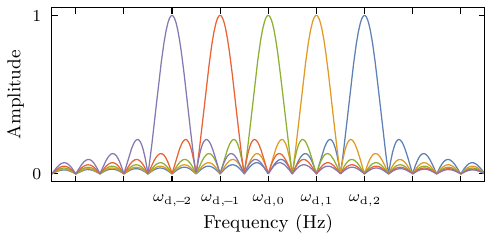}
					\put(12, 42.5){(c)}
				\end{overpic}
			\end{minipage}
		\caption{Absolute values of the normalized spectra of $s\left(t\right) \sin\left(\omega_{\mathrm{d}, k} t\right)$ for $N=5$ and $k \in \left\{-2, -1, 0, 1, 2\right\}$, with pulse length (a) $\tau = \tau_0$, (b) $\tau  = \tau_0/2$, and (c) $\tau = 2 \tau_0$. 
			In (a) and (c), each drive frequency $\omega_{\mathrm{d}, k}$ is placed at the zero-crossing point of the spectral components of all other microwaves.
		}
		\label{fig:spectrum}
	\end{figure}
	\begin{figure}[t]
\begin{minipage}[t]{1\linewidth}
				\centering
				\includegraphics[scale=1]{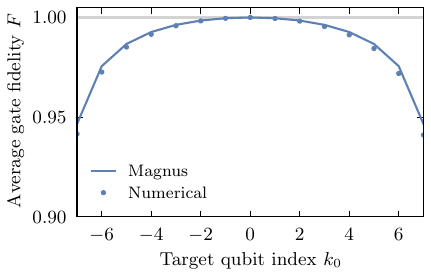}
\end{minipage}
			\begin{minipage}[t]{1\linewidth}
				\centering
				\includegraphics[scale=1]{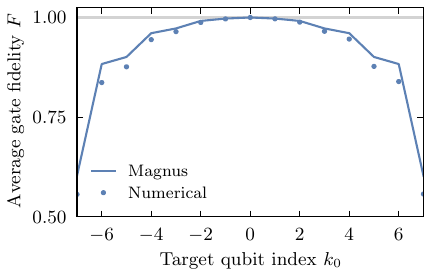}
\end{minipage}
			\begin{minipage}[t]{1\linewidth}
				\centering
				\includegraphics[scale=1]{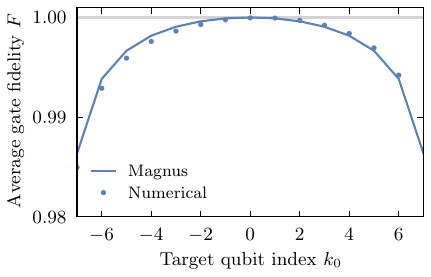}
\end{minipage}
\begin{picture}(700, 0)
			\put(58, 393){(a)}
			\put(58, 259){(b)}
			\put(58, 125){(c)}
\end{picture}
		\caption{Average gate fidelity $F\left(U_{\text{ideal}}, U_{\text{Magnus}}\left(\tau\right)\right)$ (solid lines) and $F\left(U_{\text{ideal}}, U\right)$ (dots) as a function of the target qubit index $k_0$, for three pulse lengths: (a) $\tau = \tau_0$, 
				(b) $\tau = \tau_0/2$, and 
				(c) $\tau = 2 \tau_0$.
				Parameters are set as $N = 15$, $-l = r =7$, $\omega_{\mathrm{q}, k_0} / 2\pi = 5 \,\unit{\giga\hertz}$, $\Delta/2\pi = 10 \,\unit{\mega\hertz}$, $\phi = \pi / 2$, and $\alpha = \phi/\tau$.
				}
		\label{fig:result_case_studies}
	\end{figure}
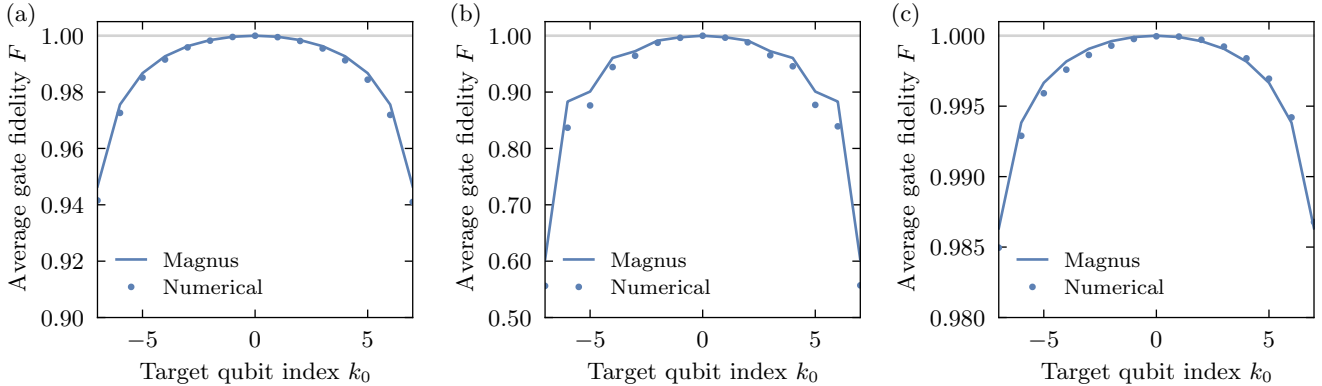

Here, we demonstrate that high fidelity can be obtained for several specific pulse length values~$\tau$. 
	Specifically, we analyze cases in which $\tau = \tau_{0}$ (Sec.~\ref{sec:tau=tau0}), $\tau = \tau_{0}/2$ (Sec.~\ref{sec:tau=tau0/2}), and $\tau \gg \tau_{0}$ as the long-pulse limit (Sec.~\ref{sec:tau_large}), where $\tau_{0} \equiv 2 \pi / \Delta$.
	
	Note that a larger frequency spacing $\Delta$ mitigates the influence of the off-resonant drive.
	A shorter pulse length $\tau$ corresponds to a wider spectral width, requiring a larger $\Delta$ to attain high fidelity; conversely, reducing $\Delta$ requires increasing $\tau$. 
	In this sense, minimizing $\tau$ for a fixed $\Delta$ is essentially equivalent to minimizing $\Delta$ for a fixed $\tau$; both increase the gate operation throughput (i.e., the number of gates executable per unit time) within a given total bandwidth. 
	For a practical implementation, $\tau$ can be adjusted by tuning the microwave controller, which is generally a more straightforward method than manipulating $\Delta$.
	Therefore, in the remainder of this paper, we focus on identifying the pulse lengths that yield higher fidelity values rather than varying $\Delta$.

	\subsection{Orthogonal case ($\tau = \tau_{0}$)} \label{sec:tau=tau0}
		We consider $\tau = \tau_0$, where the microwave signals are mutually orthogonal. 
		That is, the peak of the frequency spectrum of one microwave signal aligns with the null points of the spectra of all the other microwave signals, as illustrated in Fig.~\ref{fig:spectrum}(a).
		Such a spectrum allocation is commonly used in wireless communication systems and is known as OFDM, enabling subcarrier signals to overlap in the frequency domain without causing interference~\cite{OFDM, OFDM2009, 5G}. 
		Therefore, it is anticipated that high gate fidelity can be achieved when $\tau = \tau_0$.
		
		Notably, because $\sin\left(n \Delta \tau_0\right) = 0$ and $\sin\left(n\frac{\Delta\tau_0}{2}\right) = 0$ for all integers $n$, Eqs.~\eqref{eq:lambda_x_general},
		\eqref{eq:lambda_y_general}, and \eqref{eq:lambda_z_general} reduce to 
		\begin{align}
			\label{eq:lambda_x_0}
			\lambda_{\mathrm{x}} &= \frac{\phi}{2}, \\
			\label{eq:lambda_y_0}
			\lambda_{\mathrm{y}} &= 0, \\
\label{eq:lambda_z_0}
			\lambda_{\mathrm{z}} &= \frac{\phi^2}{8\pi} \sum_{k \in \gamma} \frac{1}{k-k_0},
		\end{align}
		when $\alpha = \phi/\tau$.
		This reduction suggests that the interference from off-resonant microwaves is suppressed, thereby achieving high gate fidelity. 
		In fact, we observe that $F\left(U_{\text{ideal}}, U_{\text{Magnus}}\left(\tau\right)\right) \sim 1$ near $k_0 = 0$ at $\tau = \tau_{0}$, as shown in Fig.~\ref{fig:result_case_studies}(a). Here, $F\left(U_{\text{ideal}}, U_{\text{Magnus}}\left(\tau\right)\right)$ is calculated using Eqs.~\eqref{eq:F_analytical} and \eqref{eq:lambda_x_0}--\eqref{eq:lambda_z_0} for $N = 15$, $-l = r =7$, and $\phi=\pi/2$.
		This result indicates that orthogonal drive frequencies provide high gate fidelity. 
		
		Note that the fidelity degrades as $|k_0|$ increases. 
		This degradation is attributed to the corresponding increase in $\left|\lambda_\mathrm{z}\right|$, which conflicts with the condition for high fidelity, which requires $\lambda_\mathrm{z}$ to be zero [see Fig.~\ref{fig:result_lambda}(a)].

	\subsection{Quasi-orthogonal case ($\tau = \tau_{0}/2$)} \label{sec:tau=tau0/2}
		\begin{figure}[tb]
				\begin{minipage}[t]{1\linewidth}
					\centering
					\begin{overpic}[scale=1]{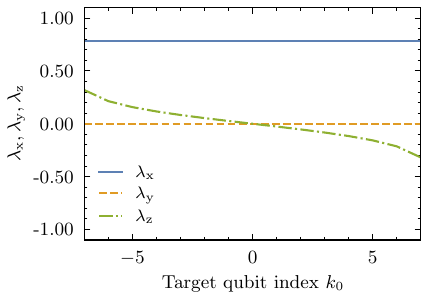}
						\put(1, 65){(a)}
					\end{overpic}
				\end{minipage}
				\begin{minipage}[t]{1\linewidth}
					\centering
					\begin{overpic}[scale=1]{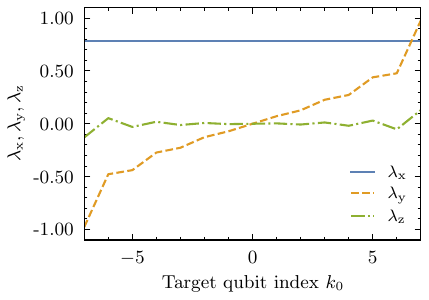}
						\put(1, 65){(b)}
					\end{overpic}
				\end{minipage}
			\caption{Rotation angles $\lambda_{\mathrm{x}}$, $\lambda_{\mathrm{y}}$, and $\lambda_{\mathrm{z}}$ as a function of the target qubit index $k_0$ for (a)~orthogonal ($\tau = \tau_{0}$) and (b)~quasi-orthogonal  ($\tau = \tau_{0}/2$) cases. 
			Parameters are set as $N = 15$, $-l = r =7$, and $\phi = \pi / 2$. 
			Results are obtained using Eqs.~\eqref{eq:lambda_x_0}--\eqref{eq:lambda_z_0} for (a) and Eqs.~\eqref{eq:lambda_x_1}--\eqref{eq:lambda_z_1} for (b). }
			\label{fig:result_lambda}
	   \end{figure}

		Next, we examine the case where $\tau = \tau_0/2$, which is half the duration required for pulse orthogonality, and illustrate the spectra of the microwave signals in Fig.~\ref{fig:spectrum}(b).
		In this case, the microwaves are no longer mutually orthogonal.
		 Equations~\eqref{eq:lambda_x_general}, \eqref{eq:lambda_y_general}, and \eqref{eq:lambda_z_general} are rewritten as follows:
		\begin{align}
			\label{eq:lambda_x_1}
			\lambda_{\mathrm{x}} &= \frac{\phi}{2}, \\
			\label{eq:lambda_y_1}
			\lambda_{\mathrm{y}} &= \frac{\phi}{2\pi} \sum_{k\in\gamma} \frac{\left(-1\right) ^{k-k_0} - 1}{k-k_0}, \\
\label{eq:lambda_z_1}
			\lambda_{\mathrm{z}} &= \frac{\phi^2}{4\pi} \sum_{k \in \gamma} \frac{\left(-1\right)^{k-k_0}}{k-k_0}, 
		\end{align}
		when $\alpha = \phi/\tau$. 
		Using Eqs.~\eqref{eq:F_analytical} and \eqref{eq:lambda_x_1}--\eqref{eq:lambda_z_1}, $F\left(U_{\text{ideal}}, U_{\text{Magnus}}\left(\tau\right)\right)$ is plotted as a function of $k_0$ in Fig.~\ref{fig:result_case_studies}(b).
		The same parameters as those used in Fig.~\ref{fig:result_case_studies}(a) are used, except for pulse length $\tau$.
		Interestingly, the average gate fidelity for $\tau = \tau_{0}/2$ [Fig.~\ref{fig:result_case_studies}(b)] is qualitatively similar to that for $\tau = \tau_{0}$ [Fig.~\ref{fig:result_case_studies}(a)] around $k_0 = 0$. 
		In other words, the system exhibits high gate fidelity.
		
		However, an increase in $\left|k_0\right|$ leads to a larger degradation in the gate fidelity than in the case where $\tau = \tau_0$.
		This degradation arises from the difference in the behavior of $\lambda_{\mathrm{y}}$ and $\lambda_{\mathrm{z}}$ between $\tau = \tau_0/2$ and $\tau = \tau_0$, as shown in Fig.~\ref{fig:result_lambda}(b).
The value of $\left|\lambda_{\mathrm{y}}\right|$ increases with $\left|k_0\right|$ [dashed line in Fig.~\ref{fig:result_lambda}(b)] at $\tau = \tau_0/2$, which is a significant factor in fidelity reduction. 
		In contrast, at $\tau = \tau_0$, an increase in $\left|\lambda_{\mathrm{z}}\right|$ [dash-dotted line in Fig.~\ref{fig:result_lambda}(a)] contributes to a reduction in fidelity.
The magnitude of $\lambda_{\mathrm{y}}$ for $\tau = \tau_{0}/2$ is clearly larger than that of $\lambda_{\mathrm{z}}$ for $\tau = \tau_{0}$, resulting in smaller gate fidelity when $\tau = \tau_{0}/2$.
		
	\subsection{Long-pulse limit ($\tau \gg \tau_{0}$)} \label{sec:tau_large}
		\begin{figure}[tb]
			\centering
			\includegraphics[scale=1]{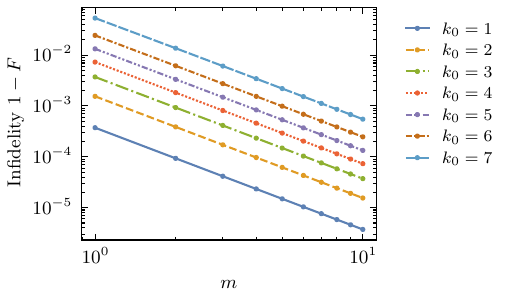}
				\caption{Infidelity $1 - F\left(U_{\text{ideal}}, U_{\text{Magnus}}\left(\tau\right)\right)$ as a function of the integer $m = \tau/\tau_0$.
				The parameters are $N = 15$, $- l = r = 7$, $\phi = \pi / 2$, and $\alpha = \phi/\tau$. 
				Since the infidelity shows similar behavior for $k_0 = \pm k$ (where $k$ is a positive integer) and the infidelity becomes zero for $k_0 = 0$, the results for $k_0 \leq 0$ are omitted here.
}
			\label{fig:result_infidelity_vs_n}
		\end{figure}
		
		Finally, we present the results for $\tau \gg \tau_0$, where
		a longer pulse length corresponds to a narrower spectral width of the drive pulse.
		Considering the orthogonal microwaves $\tau = m \tau_0$, where $m$ is a positive integer, 
		Eqs.~\eqref{eq:lambda_x_general}, \eqref{eq:lambda_y_general}, and \eqref{eq:lambda_z_general} can be simplified as follows:
		\begin{align}
			\label{eq:lambda_x_2}
			\lambda_{\mathrm{x}} &= \frac{\phi}{2}, \\
			\label{eq:lambda_y_2}
			\lambda_{\mathrm{y}} &= 0, \\
\label{eq:lambda_z_2}
			\lambda_{\mathrm{z}} &= \frac{\phi^2}{8\pi m} \sum_{k \in \gamma} \frac{1}{k-k_0}, 
		\end{align}
		when $\alpha = \phi/\tau$. 
		We now focus on $\tau = 2 \tau_0$, which corresponds to the spectra shown in Fig.~\ref{fig:spectrum}(c). Figure~\ref{fig:result_case_studies}(c) shows the gate fidelity $F\left(U_{\text{ideal}}, U_{\text{Magnus}}\left(\tau\right)\right)$ calculated using Eqs.~\eqref{eq:F_analytical} and \eqref{eq:lambda_x_2}--\eqref{eq:lambda_z_2}.
The fidelity is higher than that for $\tau = \tau_0$ [Fig.~\ref{fig:result_case_studies}(a)] and $\tau = \tau_0/2$ [Fig.~\ref{fig:result_case_studies}(b)] for the entire range of $k_0$.

		Moreover, using Eqs.~\eqref{eq:lambda_x_2}--\eqref{eq:lambda_z_2}, we plot the infidelity $1 - F\left(U_{\text{ideal}}, U_{\text{Magnus}}\left(\tau\right)\right)$ as a function of the integer $m = \tau/\tau_0$ in Fig.~\ref{fig:result_infidelity_vs_n}, 
		employing the same parameters as those used in Fig.~\ref{fig:result_case_studies}.
		Notably, $1 - F\left(U_{\text{ideal}}, U_{\text{Magnus}}\left(\tau\right)\right)$ fits well to linear functions on a log-log scale with slopes of approximately $-2.0$. 
		This indicates that the infidelity $1 - F\left(U_{\text{ideal}}, U_{\text{Magnus}}\left(\tau\right)\right)$ decreases proportionally to $\tau^{-2}$ when $\tau = m \tau_0$. 
		
		We now turn to the long-pulse limit $\tau \rightarrow \infty$, without restricting $\tau$ to $m \tau_0$. 
		Assuming $\alpha = \phi/\tau$ and constant $\Delta$, Eqs.~\eqref{eq:lambda_x_general}, \eqref{eq:lambda_y_general}, and \eqref{eq:lambda_z_general} become 
		\begin{align}
			\label{eq:lambda_x_3}
			\lambda_{\mathrm{x}} &\rightarrow \frac{\phi}{2}, \\
			\label{eq:lambda_y_3}
			\lambda_{\mathrm{y}} &\rightarrow 0, \\
			\label{eq:lambda_z_3}
			\lambda_{\mathrm{z}} &\rightarrow 0. 
		\end{align}
		Hence, we obtain 
		\begin{align}
			\label{eq:Magnus_limit}
				U_{\text{Magnus}}\left(\tau\right) \rightarrow e^{ i \frac{\phi}{2} \sigma_{\mathrm{x}, k_0}} = U_{\text{ideal}}, 
		\end{align}
		and therefore, 
		\begin{align}
		\label{eq:F_limit}
				F\left(U_{\text{ideal}}, U_{\text{Magnus}}\left(\tau\right)\right) \rightarrow 1. 
		\end{align}
		Thus, higher gate fidelity can be achieved with a longer pulse length. 
		This result is consistent with our intuition, that is, longer pulse lengths correspond to less interference.

\section{Parameter studies} \label{sec:param_study_FDM}
	\begin{figure*}[tb]
\begin{tabular}{ccc}
			\begin{minipage}[t]{0.33\linewidth}
				\centering
				\includegraphics[scale=1]{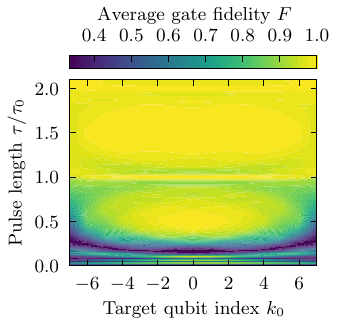}
			\end{minipage}
			\begin{minipage}[t]{0.33\linewidth}
				\centering
				\includegraphics[scale=1]{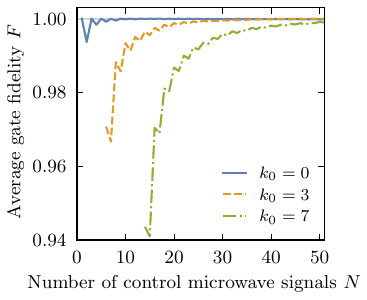}
			\end{minipage}
			\begin{minipage}[t]{0.33\linewidth}
				\centering
				\includegraphics[scale=1]{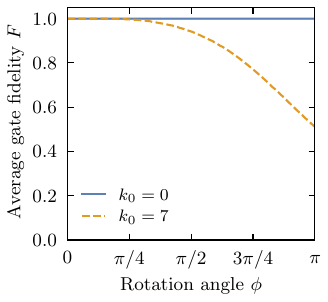}
			\end{minipage}
		\end{tabular}
		\begin{picture}(700, 0)
			\put(8, 152){(a)}
			\put(174, 152){(b)}
			\put(352, 152){(c)}
		\end{picture}
\caption{
				Average gate fidelity $F\left(U_{\text{ideal}}, U\right)$
				as a function of 
				(a) target qubit index $k_0$ and pulse length $\tau$,  
				(b) number of control microwave signals $N$, 
				and (c) rotation angle $\phi$. 
				Common parameters for (a)--(c) are $\omega_{\mathrm{q}, 0}/2\pi = 5 \,\unit{\giga\hertz}$, 
				$\Delta/2\pi = 10 \,\unit{\mega\hertz}$, and $\alpha = \phi/\tau$. 
				The other parameters are (a) $N = 15$, $-l = r =7$, and $\phi = \pi / 2$; 
				(b) $l =  -\lfloor \left(N - 1\right) / 2\rfloor$, $r =  \lfloor N / 2\rfloor$, $\phi = \pi / 2$, and $\tau = \tau_0$; 
				and (c) $N = 15$, $-l = r = 7$, and $\tau = \tau_0$. 
In (b), the fidelity is not shown in $N < 6$ for $k_0 = 3$ and in $N < 14$ for $k_0 = 7$, since the drive microwave components for those qubits do not exist in those regimes.
}
		\label{fig:result_param}
	\end{figure*}
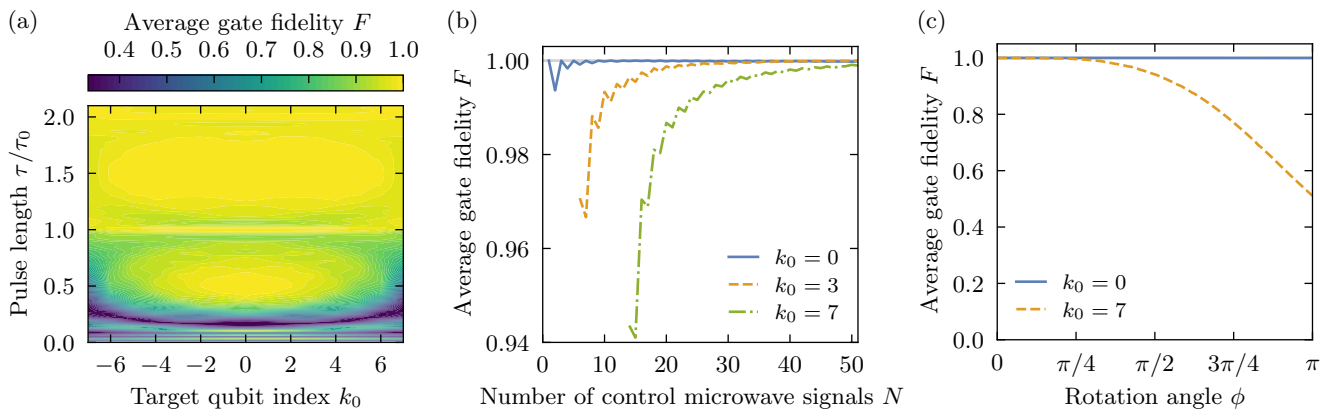
	
	Here, we numerically investigate the relationship between the fidelity $F\left(U_{\text{ideal}}, U\right)$ and the key parameters: pulse length $\tau$ (Sec.~\ref{sec:tau}), number of control microwave signals $N$, corresponding to the number of qubits (Sec.~\ref{sec:num_freq}), and rotation angle $\phi$ (Sec.~\ref{sec:phi}). 
	The time evolution of the system is obtained numerically by simulating the Schr\"{o}dinger equation using the Hamiltonian given by Eq.~\eqref{eq:H}.

	\subsection{Pulse length $\tau$} \label{sec:tau}
		We analyze the dependence of the average gate fidelity on the pulse length $\tau$ and target qubit index $k_0$ to identify suitable pulse lengths that yield high gate fidelity.
		Note again that a change in $\tau$ is equivalent to a change in $\Delta$ 
		as described in Sec.~\ref{sec:case_study_FDM}. Figure~\ref{fig:result_param}(a) shows a map of $F\left(U_{\text{ideal}}, U\right)$ for $N = 15$ qubits. Throughout the entire region $\tau/\tau_0 > 1$, the fidelity is close to unity because a large $\tau/\tau_0$ corresponds to a narrow lobe width of the spectrum and mitigates the interference from the off-resonant drive.
		In addition, at $\tau / \tau_{0} = 1$, mutually orthogonal microwave signals provide locally maximized fidelity.
		Even at $\tau / \tau_0 = 0.5$, the fidelity exhibits a local maximum.
		These results are consistent with the theoretical results presented in Sec.~\ref{sec:case_study_FDM}.

	\subsection{Number of control microwave signals $N$} \label{sec:num_freq}
We investigate the relationship between the average gate fidelity $F\left(U_{\text{ideal}}, U\right)$ and the number of control microwave signals $N$ (equal to the number of qubits).
		Figure~\ref{fig:result_param}(b) shows $F\left(U_{\text{ideal}}, U\right)$ as a function of $N$ for $\tau = \tau_0$. 
		An increase in $N$ leads to higher fidelity. 
		The effect of increasing $N$ becomes more pronounced as $k_0$ increases.
This observed fidelity enhancement is achieved by the deliberate incorporation of suitably chosen off-resonant frequency components into the control pulse. 
		This finding challenges the conventional view that off-resonant microwave components are merely detrimental to qubit control. 
		Therefore, our results align with the emerging understanding of recent research, exemplified by methods such as active leakage cancellation demonstrated in transmon qubits~\cite{ALC}.

		Notably, the fidelity curves exhibit a jagged structure.
		This behavior arises from the asymmetry in the frequency allocation of the control microwave signals: 
		$F\left(U_{\text{ideal}}, U\right)$ for an operation on the qubit with frequency $\omega_{\mathrm{d}, k_0}$ 
		tends to be higher when the microwave frequencies are symmetrically allocated around it.

	\subsection{Rotation angle $\phi$} \label{sec:phi}	
		In addition to $\tau$ and $N$, the rotation angle $\phi \propto \alpha$ associated with each pulse is a key parameter for the gate operations.
		Figure~\ref{fig:result_param}(c) shows the average gate fidelity $F\left(U_{\text{ideal}}, U\right)$ as a function of $\phi$ for $N = 15$, $\tau = \tau_0$, and $\alpha = \phi/\tau$. 
		We plot the fidelity for two different target qubit indices: $k_0 = 0$ and $k_0 = 7$.
		For $k_0 = 0$, high fidelity is maintained over the entire range of $0 \le \phi \le \pi$. 
		However, for $k_0 = 7$, the fidelity decreases rapidly as $\phi$ increases.
		Therefore, a small rotation angle $\phi$ (or equivalently a small $\alpha$) is recommended for high-fidelity gate operations.

\section{Discussion}
    The pulse length is a crucial factor for enabling dense spacing of drive frequencies while maintaining high gate fidelity. 
    From the results presented in Secs.~\ref{sec:tau=tau0},~\ref{sec:tau=tau0/2}, and~\ref{sec:tau}, we observe that high fidelity can be attained near $k_0 = 0$ at both $\tau = \tau_0$ and $\tau = \tau_0/2$. 
This is because the rotation angles are $\lambda_{\mathrm{x}} \sim \phi/2$ and $\lambda_{\mathrm{y}}, \lambda_{\mathrm{z}} \sim 0$ in each case, which is necessary to achieve high fidelity. 
    An equivalent result can be obtained for every integer multiple of $\tau_0$ and $\tau_0/2$. 
    Consequently, we conclude that $\tau$ should be chosen from the set $\left\{m\tau_0/2\right\}$, where $m$ is a positive integer, to obtain higher fidelity around $k_0 = 0$.
    Note that rectangular pulses with $\tau = m\tau_0/2$ can achieve higher gate fidelity than FDM using Gaussian pulses (see Appendix~\ref{sec:appendix:gaussian}). 
    In addition, a larger $\tau$ guarantees high fidelity because the rotation angles approach $\lambda_{\mathrm{x}} \rightarrow \phi/2$ and $\lambda_{\mathrm{y}}, \lambda_{\mathrm{z}} \rightarrow 0$ in the limit of infinite pulse duration ($\tau \rightarrow \infty$), as shown in Sec.~\ref{sec:tau_large}.

    However, the nonzero $\lambda_{\mathrm{y}}$ and $\lambda_{\mathrm{z}}$ degrade the average gate fidelity when $\left|k_0\right|$ increases.
    For example, when $\tau = \tau_0$ (Sec.~\ref{sec:tau=tau0}), $\lambda_{\mathrm{x}} = \phi/2$ and $\lambda_{\mathrm{y}} = 0$ are satisfied.
    However, $\lambda_{\mathrm{z}}$ causes a reduction in fidelity.
    A closer examination reveals that $\lambda_{\mathrm{z}}$ becomes zero when the second- and higher-order terms such as $\Omega_{2}\left(\tau\right)$ are neglected in the Magnus expansion. 
    The absence of higher-order terms underlies the OFDM scheme in wireless communication systems~\cite{OFDM, OFDM2009, 5G}, where the subcarrier frequencies are closely spaced while orthogonality is maintained, thereby avoiding intercarrier interference.  
    In contrast, quantum dynamics introduce additional complexity in the context of qubit control. 
    In particular, contributions from the noncommutativity of the time-dependent Hamiltonian become significant. 

    We plot numerical results [$F\left(U_{\text{ideal}}, U\right)$] using the dots in Figs.~\ref{fig:result_case_studies}(a)--\ref{fig:result_case_studies}(c).
    The unitary operator $U$ is calculated by numerically solving the Schr\"{o}dinger equation using Eq.~\eqref{eq:H}. 
    Consequently, the theoretical results are consistent with the numerical analysis, thereby validating the approximations underlying our theoretical analysis.

    Thus far, we have assumed uniform coupling between the drive lines and the qubits. 
    In practice, however, this assumption breaks down: 
    the microwave amplitude experienced by the $k$th qubit can differ from that of others, 
    leading to an incorrect rotation angle and reduced gate fidelity. 
    Although this nonuniform coupling can be partially compensated by adjusting $\alpha_{k}$, 
    the gate fidelity of a qubit tends to be higher when the microwave frequencies are symmetrically allocated around the qubit frequency, 
    as discussed in Sec.~\ref{sec:num_freq}.
    Such compensation therefore introduces asymmetry for the remaining qubits and degrades their fidelities, 
    necessitating a system-level optimization.

    As a final remark, we have neglected leakage to higher energy levels, 
    which can limit gate fidelity in weakly anharmonic systems such as transmons. 
    This simplification allows us to isolate and clarify the essential physics of the OFDM scheme. 
    In practical implementations, rectangular pulse envelopes can induce significant leakage for typical anharmonicities of $200$--$300 \, \mathrm{MHz}$~\cite{guide}. 
    Using raised-cosine pulses can mitigate this effect due to their smoother waveforms.
    For further suppression of leakage, the derivative removal by adiabatic gate (DRAG) technique is a natural candidate. 
    In the context of OFDM, however, leakage suppression becomes more involved because each qubit has a distinct leakage transition frequency, rendering the simultaneous optimization of DRAG across multiple qubits nontrivial. 
    A comprehensive analysis of leakage outside the computational subspace is beyond the scope of the present work and will be addressed in future work.

\section{Conclusion}
	We have analyzed an FDM-based qubit control via a single control line. 
We have focused on the pulse length, equivalently considering the spacing of the drive frequencies, to achieve high-fidelity simultaneous gate operations on multiple qubits. 
	Our results indicate that orthogonality of the drive-pulse frequencies considerably improves gate fidelity. Notably, even when the frequency spacing was reduced to half of the orthogonal condition, the system retained comparable fidelity levels. 
These results potentially broaden the design space for spectrum allocation and pulse length of microwave signals.

	Our approach shares conceptual similarities with OFDM, a technique widely used in wireless communication systems. 
	However, unlike OFDM in wireless communication systems, the noncommutativity of the time-dependent Hamiltonian in quantum systems introduces residual crosstalk. 
	Addressing such quantum effects to further enhance the gate fidelity is an important direction for future research.

\begin{acknowledgments}
	This study used QuTiP~\cite{qutip5} for numerical calculations.	
\end{acknowledgments}

\appendix
\section{Comparison with FDM using Gaussian pulses}\label{sec:appendix:gaussian}

    \begin{figure}[tb]
        \centering
\begin{minipage}[t]{\linewidth}
                \centering
                \includegraphics[scale=1]{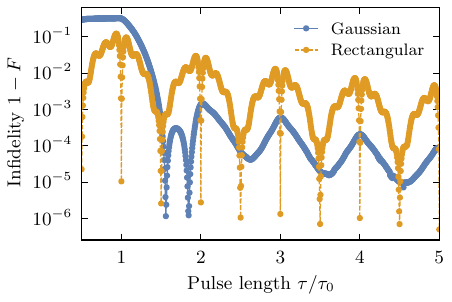}
            \end{minipage}
            \begin{minipage}[t]{\linewidth}
                \centering
                \includegraphics[scale=1]{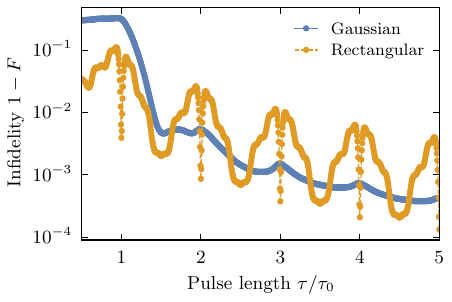}
            \end{minipage}
            \begin{minipage}[t]{\linewidth}
                \centering
                \includegraphics[scale=1]{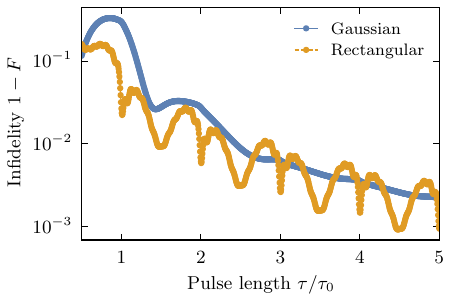}
            \end{minipage}
\begin{picture}(700, 0)
			\put(15, 438){(a)}
			\put(15, 292){(b)}
			\put(15, 146){(c)}
		\end{picture}
        \caption{Average gate infidelity $1 - F\left(U_{\text{ideal}}, U\right)$ as a function of the pulse length $\tau/\tau_0$ for Gaussian pulses and rectangular pulses with (a) $k_0 = 0$, (b) $k_0 = 3$, and (c) $k_0 = 6$. 
        Parameters are set as $N = 15$, $-l = r =7$, $\omega_{\mathrm{q}, k_0} / 2\pi = 5 \,\mathrm{GHz}$, $\Delta/2\pi = 10 \,\mathrm{MHz}$, $\phi = \pi / 2$, and $\alpha = \phi/\tau$. 
        The evolution operator $U$ is obtained numerically from the Hamiltonian in Eq.~\eqref{eq:H}. 
        }
        \label{fig:tau_envelope}
    \end{figure}

    We compare the infidelity for rectangular pulses under orthogonal and quasi-orthogonal conditions with that for Gaussian pulses. 
    Figures~\ref{fig:tau_envelope}(a)--\ref{fig:tau_envelope}(c) show the average gate infidelity $1-F\left(U_{\text{ideal}}, U\right)$ as a function of the pulse length $\tau/\tau_0$ for rectangular pulses and Gaussian pulses with $k_0=0$, $3$, and $6$. 
Here, the evolution operator $U$ is obtained numerically from the Hamiltonian in Eq.~\eqref{eq:H}. 
    For Gaussian pulses, we use the envelope given by~\cite{analytic_control_methods}
    \begin{align}
        s\left(t\right) = \phi \frac{\exp \left[ - \frac{\left( t - \tau/2 \right)^2}{2\sigma^2} \right] - \exp\left[-\frac{\tau^2}{8\sigma ^2}\right]}{\sqrt{2\pi\sigma^2} \mathrm{erf}\left[\frac{\tau}{\sqrt{8}\sigma}\right] - \tau \exp\left[-\frac{\tau^2}{8\sigma^2}\right]}, 
    \end{align}
    where $\sigma$ is the standard deviation, and we set $\sigma = \tau/4$.
    In Figs.~\ref{fig:tau_envelope}(a)--\ref{fig:tau_envelope}(c), 
    the infidelity for rectangular pulses under orthogonal and quasi-orthogonal conditions ($\tau/\tau_0 = 0.5, 1, 1.5, 2, \dots$) is lower than that for Gaussian pulses with the same or larger values of $\tau$. 
    An exception is observed in Fig.~\ref{fig:tau_envelope}(a), where Gaussian pulses yield a lower infidelity in limited regions within $1.5 < \tau/\tau_0 < 2.0$.
    These results indicate that, compared with Gaussian pulses, our approach enables more efficient utilization of the available bandwidth, analogous to OFDM in wireless communication systems.

\bibliography{bib/quantum_algorithm, bib/PQC, bib/reduce_control_line, bib/control, bib/others, bib/quantum_device, bib/FTQC}

\end{document}